\documentclass[10pt,conference]{IEEEtran}
\IEEEoverridecommandlockouts
\usepackage{cite}
\usepackage{amsmath,amssymb,amsfonts}
\usepackage{algorithmic}
\usepackage{graphicx}
\usepackage{subcaption}
\usepackage{textcomp}
\usepackage{xcolor}
\usepackage{tcolorbox}
\usepackage{multirow}
\usepackage{adjustbox}
\usepackage[switch]{lineno}
\usepackage{colortbl}
\definecolor{mycolor}{rgb}{0.8, 0.8, 0.2} 
\usepackage[hidelinks]{hyperref}

\usepackage{titlesec}
\titlespacing*{\section}{0pt}{0.5\baselineskip}{0.5\baselineskip}  
\titlespacing*{\subsection}{0pt}{0.5\baselineskip}{0.5\baselineskip}

\def\BibTeX{{\rm B\kern-.05em{\sc i\kern-.025em b}\kern-.08em
    T\kern-.1667em\lower.7ex\hbox{E}\kern-.125emX}}
\begin{document}
\title{Architectural Transformations and Emerging Verification Demands in AI-Enabled Cyber-Physical Systems}

\author{\IEEEauthorblockN{Hadiza Umar Yusuf}
\IEEEauthorblockA{
\textit{University of Michigan-Dearborn, USA}\\
hyusuf@umich.edu}
\and
\IEEEauthorblockN{Khouloud Gaaloul}
\IEEEauthorblockA{
\textit{University of Michigan-Dearborn, USA}\\
kgaaloul@umich.edu}
}

\maketitle

\begin{abstract}
In the world of Cyber-Physical Systems (CPS), a captivating real-time fusion occurs where digital technology meets the physical world. This synergy has been significantly transformed by the integration of artificial intelligence (AI), a move that, while dramatically enhancing system adaptability, also introduces a layer of complexity that impacts CPS control optimization, and reliability. Despite advancements in AI integration, a significant gap remains in understanding how this shift affects CPS architecture, operational complexity, and verification practices. This paper addresses this gap by investigating both the static and dynamic architectural distinctions between AI-driven and traditional control models designed in Simulink, as well as their respective implications for system verification. Our analysis examines two distinct versions of CPS models; those built with traditional controllers, such as Model Predictive Control (MPC) and Proportional-Integral-Derivative (PID) control, and those using AI-driven models, specifically Deep Reinforcement Learning (DRL). We focus, at a granular level, on atomic block composition, connectivity patterns, and path complexity to investigate divergences between these models. Furthermore, we evaluate the effectiveness of standard CPS verification approach when applied to AI-driven models, in comparison to traditional models. Our results highlight a shift towards discrete and logic-driven design, and show that on average, AI-driven models exhibit 25.7\% increase in core functionality blocks and 20.5\% increase in connectivity, improving adaptability but also imposing increased computational demands, potential reliability concerns, and raising challenges for verification processes. This work highlights the need for guided CPS control architecture design and adaptive verification practices to address the increasingly intelligent and interconnected systems. 
\end{abstract}

\begin{IEEEkeywords}
AI-enabled Cyber-Physical Systems, Control systems, Deep Reinforcement
Learning, Simulink Models 
\end{IEEEkeywords}

\section{Introduction}
\label{sec:introduction}
Cyber-Physical Systems (CPS)~\cite{baheti2011cyber,derler2011modeling,shi2011survey} have gained significant attention over recent years, with increasing research and industrial applications focused on its potential to address modern social and economic challenges and revolutionize various sectors. CPS are intricate systems that integrate physical processes with computational elements, enabling data processing, decision-making, control, optimization, and real-time monitoring~\cite{lee2008cyber,liu2017review}. The rapid integration of Artificial Intelligence (AI) within CPS~\cite{khalil2022literature,willard2020integrating,rai2020driven}—referred to as AI-CPS—has further amplified this transformative potential, allowing CPS to perform more adaptive operations. However, AI integration has introduced unique challenges due to the non-linear dynamics and high-dimensional, continuous state and action spaces of AI components. This complexity makes traditional design and verification tools less effective, especially in optimizing control operations in CPS. We are currently witnessing the replacement of traditional controllers~\cite{Okasha2022DesignAE, Varma2020TrajectoryTO, Dani2017PerformanceEO, Patra2017AdaptiveCM,Nugraha2022BrakeCC,Ferrari2019AdvancedCS} 
in CPS such as Model Predictive Control (MPC), Proportional-Integral-Derivative (PID) Control, and Linear Quadratic Regulator (LQR), with deep neural networks Deep Reinforcement Learning (DRL)~\cite{ lee2022deep,li2023deep,pu2023deep} to shift towards AI-enabled CPS systems. 

Recent studies comparing traditional PID control, with DRL in CPS~\cite{song2022cyber,lyu2023autorepair} reveal a lack of comparative analysis of AI-driven systems to their traditional counterparts. Although there is an increasing demand in many areas for integrating AI into CPS, this gap presents significant implications that extend beyond performance impacts; the transition to AI-CPS architectures fundamentally affects the entire development process, including activities such as design modeling, verification and validation. Notably, CPS for the automotive industry must comply with the ISO 21448 Safety of the Intended Functionality (SOTIF) standards~\cite{no202221448}, which mandates road-vehicles safety standard. However, AI-enabled CPS has become substantially more complex to verify. Consequently, existing verification methods, though effective for traditional CPS, often fall short in identifying and mitigating faults within AI-driven CPS. Thus, a thorough understanding of the static and dynamic architectural distinctions introduced by AI integration and their impact on verification practices~\cite{xie2023mosaic,munir2023artificial} is essential to advancing the adaptability, stability and reliability of AI-enabled CPS.

\begin{figure}[t]
\centerline{\includegraphics[width=0.4\textwidth]{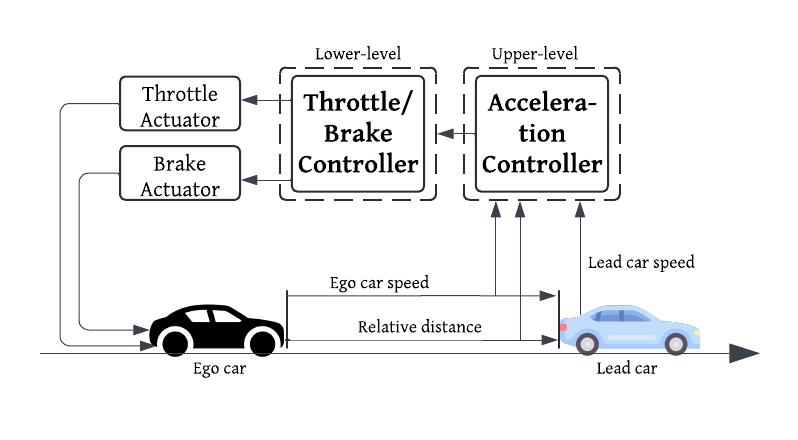}}
\caption{Control Framework of Adaptive Cruise Control System.}
\label{fig:acc}
\end{figure}

To motivate our work, we present a model of the Adaptive Cruise Control (ACC) control system. This system was published by Mathworks~\cite{pananurak2009adaptive}. This system simulates an ego car and a lead car operating in a controlled environment, where the goal is to maintain a safe distance between the two vehicles throughout the simulation. The main component of ACC is the controller that adjusts the ego car’s velocity to keep the relative distance above a desired safety threshold. The control subsystem of the ACC model is implemented in Simulink using a traditional Model Predictive Control (MPC)~\cite{afram2014theory}. The ACC model comprises atomic blocks where signal inputs, including the relative distance to the lead car and the relative velocity, flow through the system to produce these control outputs. In a recent study~\cite{song2022cyber}, the controller was modeled in a different version using a DRL controller in an attempt to shift the control from a traditional to AI-enabled construct. In the traditional MPC setup, the controller generates a control command at each time step by predicting both vehicles' motions within a finite time horizon. MPC relies on pre-collected labeled datasets to track the user-set cruising velocity while ensuring a safe following distance from the lead car. The DRL controller, on the other hand, incrementally learns the control method through continuous interaction with the environment, rather than relying on this type of data, by observing and responding to changes within the system environment. This involves training an agent within the ACC environment, which iteratively improves its strategy by continually exploring actions and updating its policy.

The ACC example highlights several key challenges that arise in transitioning from traditional to AI-driven controllers within CPS, challenges that are central to this study. First, the design of the DRL components within the model introduces complex dependencies and additional atomic blocks that significantly differ from those in the traditional MPC controller. This architectural shift necessitates a deeper understanding of the structural impact of AI integration. Second, the dynamic nature of AI-driven control alters the dynamic characteristics of CPS, affecting execution paths, connections, and hierarchical organization of the entire model. This change in the dynamic flow requires a reevaluation of the adaptability versus complexity of AI-enabled CPS in varying operational conditions. Lastly, these architectural differences complicate the CPS verification process. Traditional verification methods, while effective in deterministic settings, often struggle to capture the non-deterministic, high-dimensional behaviors introduced by AI-driven models, highlighting the need for adapted verification approaches to maintain system reliability. 

This leaves engineers with difficult decision-making as they navigate the trade-offs of adapting existing systems to support AI’s advanced capabilities, a task that requires balancing the potential benefits of AI like adaptability and flexibility, with new dependencies and interconnections that impact system complexity, scalability and integration. Additionally, traditional verification methods are often inadequate for ensuring safety and reliability in AI-enabled environments, pushing engineers to seek new verification strategies and tools. These challenges highlight the need for a systematic approach to understanding AI’s impact on CPS and to guiding engineers in making informed decisions about AI integration.

The example motivated us to investigate the trends of transitioning from traditional controllers towards AI-driven controllers in CPS at a granular level and identify their implications. Following this motivation, we outline the contributions of this study across three core phases:
\begin{itemize}
    \item  We conduct a structural composition analysis to categorize diverse CPS models and identify differences in atomic block types between AI-driven and traditional control models. Our analysis provides insights into block categories that are distinctive to AI-driven design.
    \item We perform a dynamic flow analysis, to compare the dynamic flow distinctions between AI-driven and traditional CPS models. We examine the execution paths characteristics and components' connections to uncovers the dynamic challenges introduced by AI-driven models.
    \item We evaluate the effectiveness of existing CPS verification practices in the context of AI-driven systems. We draw insights into the impact of AI integration on system reliability and the adaptability challenges of CPS verification.
\end{itemize}

To the best of our knowledge, this is the first study to conduct a comparative analysis between traditional and AI-driven CPS models and their implications on verification practices. Our findings highlight significant research opportunities in AI-enabled CPS to address the increasing demands in industry.

\textbf{Structure.} Section~\ref{sec:background} introduces the design and verification of AI-enabled CPS as our major context of this paper.
Section~\ref{sec:approach} outlines our approach to analyze and evaluate the architectural transformations in AI-enabled CPS models and their implications. Section~\ref{sec:evaluation} formalizes the research questions, describes the experimental setup and analyzes of the evaluation results. Section~\ref{sec:treats} discusses the threats to validity.
Section~\ref{sec:related} compares our work to the related work and Section~\ref{sec:conclusion} concludes the paper. This paper contains the complete research corresponding to our short paper~\cite{11030007}, including all technical details and supplementary analyses.

\section{Background}
\label{sec:background}

\begin{figure}[t]
    \centering
    \begin{subfigure}[b]{0.23\textwidth}
        \centering
        \includegraphics[width=\textwidth]{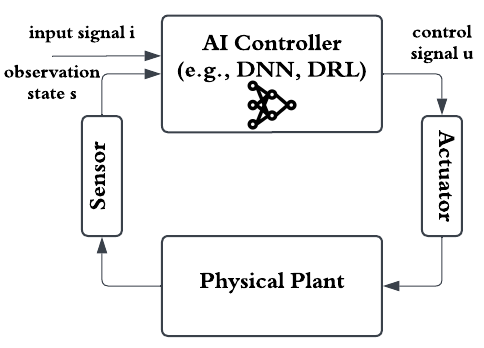}
        \caption{AI Controller}
        \label{fig:cpsai}
    \end{subfigure}
    \begin{subfigure}[b]{0.23\textwidth}
        \centering
        \includegraphics[width=\textwidth]{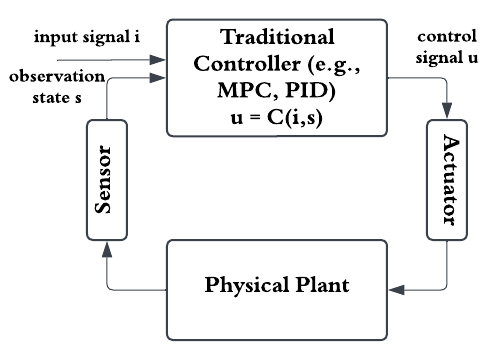}
        \caption{Traditional Controller}
        \label{fig:cpstraditional}
    \end{subfigure}
    \caption{CPS workflow with AI vs. Traditional Controller}
    \label{fig:cpsaiandt}
\end{figure}

The development of CPS has long relied on Model-Based Design (MBD)~\cite{nicolescu2018model}, which enables early-stage design, simulation, and verification. This design approach is widely adopted in fields such as robotics, aerospace, and automotive engineering~\cite{mavridou2020ten,campbell2010modeling,koenig2004design}. It consists of creating high-level, abstract models that guide the entire development process. Among functional modeling tools, the Simulink/Stateflow toolset~\cite{MathWorks2023} is well-known for designing complex systems, offering extensive libraries and domain-specific components for more robust build, test, and optimization of the system. 

Historically, CPS can be seen as a transdisciplinary domain because it integrates knowledge from multiple fields such as engineering, automation, and computer science, to create systems. The integration of AI models~\cite{baheti2011cyber,jazdi2014cyber,derler2011modeling,shi2011survey,monostori2016cyber,rajkumar2010cyber,bucsoniu2018reinforcement,radanliev2021artificial} has advanced CPS by enabling large-scale, adaptive systems capable of handling complex tasks. Leveraging deep learning and reinforcement learning~\cite{bucsoniu2018reinforcement}, AI-driven models learn from system behavior and adapt to evolving conditions. These models support high computational demands, enable adaptive control strategies, and facilitate real-time optimization, particularly in dynamic environments where traditional methods, such as those using Markov Decision Processes (MDPs)~\cite{altman2021constrained}, may fall short. 

Figure~\ref{fig:cpsaiandt} illustrates the workflow of AI-driven and traditional control process in CPS, including the physical plant and the controller. The control process, as seen in our ACC example, relies on a continuous feedback loop between system components and the external environment. For example ACC uses signals, such as system state $y$, control decision $u$, and external input $i$, as information channels within the model. Sensors and actuators enable data transmission between the physical plant (representing vehicle dynamics) and the controller (responsible for regulating vehicle speed and following distance).

Traditional control, depicted in Figure~\ref{fig:cpstraditional}, often relies on a feedback-based decision-making process, where the controller requires a known model of system dynamics. In contrast, as shown in Figure~\ref{fig:cpsai}, DRL operates without an explicit model of the system. Instead, it uses DNNs to approximate control policies and value functions, which helps it to solve problems with non-linear and stochastic dynamics. While DRL can achieved remarkable results, its "black-box" nature and reliance on deep networks adds complexity. Therefore, control engineers need to carefully evaluate its suitability for their specific CPS tasks, especially when alternative methods may offer greater stability and interpretability. 

There are several areas where AI can enhance CPS~\cite{schoning2022ai, electronics12163489}. In process modeling, AI -particularly multilayer feed-forward networks- can empirically model physical processes based on recorded data, reducing the need for iterative physical modeling. AI can be used for parameter tuning to optimize controller parameters by either providing static parameters based on typical scenarios or dynamically adjusting them in real-time through artificial neural networks (ANNs)~\cite{grossi2007introduction,yang2008artificial}. AI can further replace traditional controllers with ANNs to enable more effective interpretation of sensor feedback and, hence, improved execution of control actions. There is need for further research to improve the reliability of AI integration in CPS, though.


\section{Approach}
\label{sec:approach}
\begin{figure}
\centering
        \includegraphics[width=0.5\textwidth]{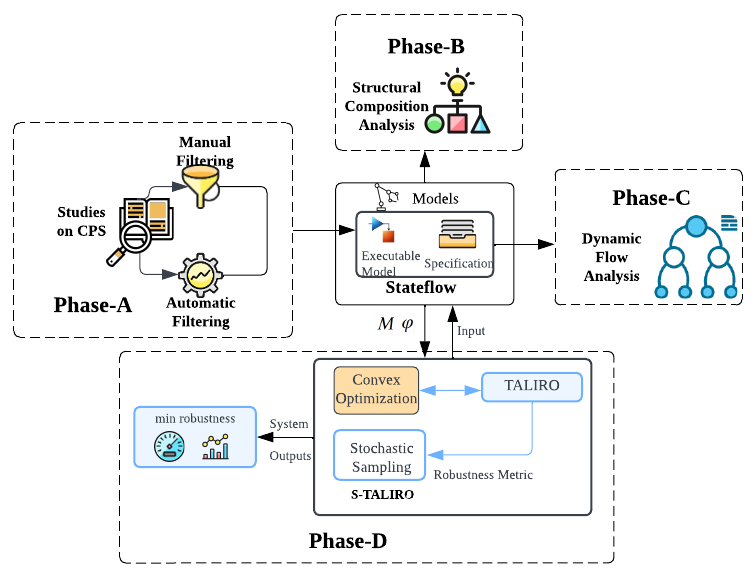}
        \setlength{\abovecaptionskip}{0pt} 
    \caption{Overview of our Multi-Method Approach.}
    \label{fig:approach}
\end{figure}

We propose a multi-method framework designed to compare the architectural characteristics of AI-driven versus traditional CPS models, and assess their impact on verification processes. As illustrated in Figure~\ref{fig:approach}, this framework follows a structured workflow to ensure effective analysis and evaluation.

 \subsection{Phase-A: Model Collection and Filtering} 
 \label{sec:step1} 
 We mine CPS models from open-source repositories, by applying a combination of manual and automated filtering techniques to identify high-quality, relevant models. The manual filtering involves expert assessment based on criteria specific to CPS design, while automatic filtering leverages algorithms to efficiently process and sift through large datasets. This dual approach ensures that the selected models align with our focus on both AI-driven and traditional CPS architectures. we consider key resources: (1) MATLAB control-related toolboxes relevant to CPS modeling and control~\cite{modelpredictivecontrol}, (2) a set of studies on cyber-physical systems, artificial intelligence, and software engineering research~\cite{zhang2020hybrid,zhang2022falsifai,song2022cyber,lyu2023autorepair,nejati2019evaluating,huang2019reachnn,tran2020nnv,althoff2015introduction,xiang2018output,song2023mathtt,xie2023mosaic}, and (3) outcomes from two workshops that gathered CPS benchmarks and held CPS verification competitions~\cite{ernst2022arch,ernst2021arch,johnson2021arch}. The filtering process has resulted in a benchmark of $8$ industry-level systems, shown in Table~\ref{tab:cpsmodels}. Each system is implemented using two control variants: a traditional PID or MPC controller and an AI-driven DRL controller. The selected models, sourced from prior research case studies~\cite{song2022cyber}, have been validated to ensure adherence to control logic and functional requirements. Our benchmark covers a diverse range of CPS applications—automotive, focusing on vehicle control and assistance, industrial automation, covering energy management, chemical processes and renewable energy, and aerospace for advanced control in rocket landing. We consider this representative benchmark to analyze CPS architecture and system dynamics across distinct domains.

\subsection{Phase-B: Structural Composition Analysis}
 
\begin{table*}
\centering
\caption{Characteristics of our Case Study Systems}
\label{tab:cpsmodels}
\begin{tabular}{p{0.04\textwidth} | p{0.17\textwidth} | p{0.5\textwidth} | p{0.15\textwidth}}
\hline
\textbf{ID} & \textbf{System Name} & \textbf{Description} & \textbf{Field} \\
\hline
ACC & Adaptive Cruise Control & A driving assistant that maintains the safety distance between cars. & Automotive \\
\hline
AFC & Abstract Fuel Control & A fuel control system for an automotive powertrain that maintains the optimal air-to-fuel ratio by adjusting the intake gas rate to the cylinder. & Automotive \\
\hline
SC & Steam Condenser & A dynamic condenser model based on energy balance and cooling water mass balance, controlled in feedback. & Energy/Power Systems \\
\hline
WT & Wind Turbine & A simplified wind turbine model, relatively large with a long time horizon (630). & Renewable Energy \\
\hline
LKA & Lane Keeping Assistant & A system that maintains the car’s trajectory along the centerline of the lanes on the road by adjusting the car's front steering angle. & Automotive \\
\hline
LR & Rocket Landing Control System & A nonlinear MPC for generating an optimal, safe landing path for a rocket at a target position. & Aerospace \\
\hline
APV & Automatic Parking Valet & A system that tracks a reference trajectory for a parking valet. & Automotive \\
\hline
CSTR & Exothermic Chemical Reactor & A chemical control system that ensures the reagent concentration in the exit stream is maintained at its desired setpoint. & Chemical Engineering \\
\hline
\end{tabular}
\end{table*}

In this phase, we analyze the structural characteristics of AI-driven CPS models in comparison to traditional CPS models. This involves identifying key differences in the types of atomic blocks and their respective categories across both model architectures. We start by tracing the atomic blocks in each of our case study system models and categorizing them by their underlying types. Table~\ref{tab:matlablibrary} provides an example list of atomic block types organized by category, representing core functional elements within CPS models. The full list of block types per category is available in our replication package~\cite{figshare2024}. For example, the \textit{Continuous} category includes fundamental control elements such as the \textit{PID Controller} and \textit{Integrator}, which are essential for managing dynamic responses in CPS models. Similarly, the \textit{Logic and Bit Operations} category involves blocks such as logic operators and relational operators, which are essential for decision-making processes and conditional logic within control architectures. We introduce our \textit{catalog of Simulink block categories}, as outlined in Table~\ref{tab:matlablibrary}, which includes a total of $8$ categories for modeling central elements to system functionality, including control logic, decision-making, data processing, and actuation. We refer to the blocks within these categories as ``relevant blocks''. Blocks outside these categories are considered ``irrelevant'', which include those used for signal attributes (e.g., data type conversion, data type duplicate, and signal specification) and sinks (e.g., scope, terminator, and display). To identify architectural distinctions between AI-driven and traditional models, we analyze the categories in our catalog that predominantly characterize AI-driven models, contrasting them with those commonly found in traditional models. This allows us to draw insights into the relationships between model constructs (i.e., AI-driven versus traditional) and their associated block categories. This phase reveals structural trends that highlight the architectural shifts introduced by AI integration in CPS.

\subsection{Phase-C: Dynamic Flow Analysis}
\label{sec:step3}
In this phase, we analyze the dynamic flow introduced by the AI integration in CPS. For each CPS model in our benchmark, we generate a control flow graph that visually represents the connectivity and dependencies among various components and atomic blocks within each model. We compare the connection frequency between AI-driven and traditional models, and we gain insights into how additional dependencies in AI-driven models may contribute and increased complexity within CPS models. Each flow graph consists of nodes and connections that represent the interactions within a system model. 

\begin{figure}
\centering
        \includegraphics[width=0.3\textwidth]{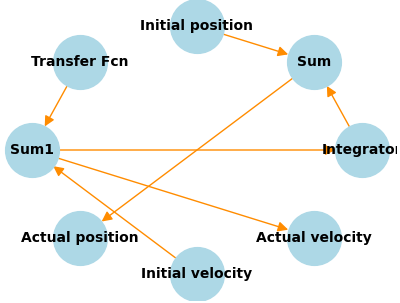}
        \setlength{\abovecaptionskip}{0pt} 
    \caption{A Simple Flow Graph form the ACC Model}
    \label{fig:samplecfg}
\end{figure}

Figure~\ref{fig:samplecfg} presents an example of a control flow graph representing a small portion of the connections within the ACC model. The nodes in the figure, labeled around the circle, refer to the block types used to model ACC. For example, ``Integrator'' is a Simulink block within the ACC model that accumulates the input signal over time to provide an integrated output. The edges between nodes represent the connections between these components, i.e, data flow, control dependencies, or interactions. A higher number of nodes suggests a more complex control structure within the system, while thicker edges indicate stronger or more frequent interactions between specific blocks.

We define and apply three \textit{dynamic flow metrics} to quantitatively assess the connectivity and dependencies in AI-driven models compared to the traditional counterpart. Higher values suggest a higher likelihood of instability, as additional dependencies can lead to unpredictable behavior when managing diverse scenarios. 
In the following, we present the \textit{dynamic flow metrics} we use for evaluating the dynamic flows in our CPS models:

\textbf{Block Count (BC)} represents the total number of atomic blocks within each model, organized by their respective Simulink block categories. Let \emph{n} denote the total number of distinct atomic block types across categories within the model, and let \emph{$b_i$} represent the occurrence of each specific block type within subsystems. The Block Count, denoted as \emph{BC}, is computed as the sum of occurrences of either relevant or total blocks, as defined by specific selection criteria, and is expressed as:

\begin{equation}
    BC = \sum_{i=1}^{n} b_i
\end{equation}

\textbf{Connection Count (CC)} quantifies the inter-connectivity between system blocks. Let \emph{$e_j$} represent each individual connection (edge) between blocks within the model, and let \emph{m} denote the total number of such connections. The Connection Count, denoted as \emph{CC}, is calculated as the sum of of all connections between either relevant or total blocks, as defined by specific selection criteria, and is given by:

\begin{equation}
    CC = \sum_{j=1}^{m} e_j
\end{equation}

\textbf{Hierarchical Depth (HD)} reflects the levels of branching within a given model, representing how many layers or subsystems are embedded within the design. Greater hierarchical depth indicates a more complex, layered structure. Let \emph{HD} denote the Hierarchical Depth of the model, which represents the depth or level of nesting of subsystems within the model. Hierarchical Depth is calculated by traversing the model's nested subsystems, from the top-level system to the deepest nested subsystem. Given the depth level of each subsystem \emph{$d_i$}, where the top-level system has $d = 1$ and each subsequent nested subsystem increases the depth by $1$, \emph{HD} is defined as:
\begin{equation}
    HD = \max(d_i)
\end{equation}
We analyze the \textit{dynamic flow metrics} within our benchmark to gain a detailed view of the dependencies and interconnections specific to AI-driven and traditional control models, and we evaluate the extent to which they differ.


\subsection{Phase-D: Implications on CPS verification}
In the final stage, we assess the implications of the identified architectural differences on the CPS verification process. We evaluate the fault-detection capabilities of standard falsification in both AI-driven and traditional models, and we highlight areas where it may require adaptation to address the complexities introduced by AI integration. Falsification-based testing~\cite{abbas2015test,pizer2016falsification,perez2022test,snyder1981testing,hoxha2014towards,gaaloul2020mining,gaaloul2021combining,nejati2019evaluating} is a widely used technique for identifying model behaviors that violate system specifications, by executing the system with a range of sampled test inputs. \emph{S-TaliRo} falsification tool~\cite{staliro, annpureddy2011staliro} stands out as a widely known modular software tool for verification and testing of CPS modeled in Simulink, having demonstrated success in several ARCH-COMP~\cite{ernst2022arch} competitions with multiple falsifiers~\cite{menghi2020approximation,corso2021survey, donze2010breach} across various CPS models. To assess the implications of AI integration in CPS, we conduct two structured experiments using \emph{S-TaliRo}, evaluating its fault-detection effectiveness across a selection of CPS models, as detailed in Table~\ref{tab:cpsmodels}. These experiments assess \emph{S-TaliRo}’s capability to detect requirement violations in both AI-driven and traditional models. First, we evaluate how effectively \emph{S-TaliRo} identifies violations in AI-driven models compared to traditional models in our case studies, with the AI-driven models configured using the Deep Deterministic Policy Gradient (DDPG)~\cite{qiu2019deep} policy. Second, we examine \emph{S-TaliRo}’s robustness and efficiency in detecting requirement violations under four distinct AI policies: DDPG, Twin-Delayed Deep Deterministic Policy Gradient~\cite{dankwa2019twin} (TD3), Actor-Critic~\cite{su2017sample} (A2C), and Proximal Policy Optimization~\cite{schulman2017proximal} (PPO). This comparative approach provides insights into the impact of different AI policies on verification performance, identifying challenges and potential adaptations necessary for effective verification of AI-enabled CPS.


\section{Evaluation}
\label{sec:evaluation}

\begin{table*}
    \centering
    \caption{Our Catalog of Simulink Block Categories and Associated Block Types (See Simulink Block Library~\cite{MathWorks2024})}
    \label{tab:matlablibrary}
    \centering
    \begin{tabular}{|p{0.03\linewidth}|p{0.16\linewidth}|p{0.7\linewidth}|}
    \hline
    \textbf{ID} & \textbf{Categories} & \textbf{Block Types} \\
    \hline
    C1 & Continuous & Derivative, Transfer Fcn, Integrator, Transport Delay, State-Space, Descriptor State-Space, Entity Transport Delay, First Order Hold, PID Controller, Second-Order Integrator,  Variable Time Delay, e.t.c.\\
    \hline
    C2 & Discontinuities & Saturation, Dead Zone, Quantizer, Rate Limiter, Backlash, Coulomb and Viscous Friction, Dead Zone Dynamic, Hit Crossing, Relay, Variable Pulse Generator, Dead Zone Dynamic, PWM, e.t.c.\\
    \hline
    C3 & Discrete & Discrete-Time integrators, Discrete Derivative , Discrete Filter, Discrete FIR Filter, Discrete PID Controller, Discrete State-Space, Discrete Transfer Fcn, Discrete Zero-Pole, Discrete-Time Integrator, Memory, e.t.c.\\
    \hline
    C4 & Logic and Bit Operations & Logic Operators, Relational Operators, Shift Arithmetic, Interval Test, Compare to Zero, Compare to Constant, Combinatorial Logic, Detect Change, Detect Decrease, Detect Fall Negative, Detect Fall Nonpositive, e.t.c.\\
    \hline
    C5 & Math Operations & Algebraic/non-Algebraic Operations, Algebraic Constraint, Gain, Assignment, Bias, Complex to Magnitude-Angle, Complex to Real-Imag, Find Nonzero Elements, Reshape, Rounding Function, Sign, e.t.c. \\
    \hline
    C6 & Ports \& Subsystems & Switch Case, Enable, Functtion Element, If, Inport, Outport, Model Trigger, Unit System Configuration, 	Template subsystem containing Subsystem blocks as variant choices While Iterator Subsystem, e.t.c. \\
    \hline
    C7 & Sources & Random Number, Band-Limited White Noise,	Chirp Signal, Clock, Constant, Counter Free-Running, Digital Clock, Enumerated Constant, From File, From Spreadsheet, From Workspace, Ground, In Bus Element, e.t.c.	\\
    \hline
    C8 & User-Defined Functions & Fcn, Interpreted MATLAB Function, MATLAB Function, MATLAB System, Reinitialize Function, Reset Function, S-Function, S-Function Builder, Simulink Function, Function Caller, Terminate Function, e.t.c.\\
    \hline
    \end{tabular}
\end{table*}

We implemented a multi-method approach, which combines an in-depth architectural analysis of CPS models, followed by a systematic evaluation to assess the impacts of AI integration in CPS on the verification process. In this section, we evaluate our framework to address the following research questions:

\textbf{RQ1:} \textit{What structural and architectural differences distinguish AI-driven from traditional control in CPS models?} This question seeks to understand how AI-driven control structures differ from traditional control methods within CPS architectures by conducting a structural composition analysis of CPS models. We focus on categorizing atomic block types and identifying unique block categories prevalent in AI-driven models and we provide insights into the architectural shifts introduced by AI integration and their impact on the complexity of control design in AI-enabled CPS.

\textbf{RQ2:} \textit{How do the dynamic flow characteristics of AI-driven control models differ from traditional control models in Cyber-Physical Systems?} This question investigates the dynamic distinctions between AI-driven and traditional control models in CPS through dynamic flow analysis. To answer this question, we examine execution paths and inter-component connections, and we analyze control flow dependencies, and hierarchical structure to identify the adaptability and complexity challenges that AI-driven models introduce.

\textbf{RQ3:} \textit{How does AI integration in CPS models impact the effectiveness of verification processes?} This question evaluates the effects of AI integration on CPS verification, particularly in detecting faults and verifying functional requirements. We test the fault-detection capabilities of standard verification, and we identify specific challenges that arise in verifying AI-driven CPS models, including the need for adapted fault-detection and verification techniques to address the increased complexity that AI integration brings to CPS.


\subsection{Experimental Settings}
Following the model collection and filtering phase described in Section~\ref{sec:step1}, we select a benchmark of eight $(8)$ Simulink models (detailed in Table~\ref{tab:cpsmodels}) for comparative analysis of architectural composition. To ensure fair comparison, we manually perform a sanity check to confirm that both AI-driven and traditional model constructs meet the objectives of the system specification. Through the experiments\footnote{using Windows 11 Pro, Intel Core i7-1255U, 16 GB RAM.
}, we verified simulation compatibility by ensuring that all models could be run under identical conditions with standardized simulation parameters (e.g., solver settings, step size). This process guarantees that observed differences in structural metrics reflect inherent model characteristics rather than variations in simulation setups.
Each system in our benchmark is modeled with two controller versions-one AI-driven and one traditional.  The AI-driven models  across all case study systems use a DRL controller. For the traditional models, five models (i.e., ACC, APV, LKA, CSTR, and LR) use an MPC controller; AFC has a two-part control system consisting of (1) a PI controller and (2) a feed-forward controller; and WT and SC utilize a PID controller.


\subsection{RQ1}

To address \textit{RQ1}, we conduct a comparative analysis across AI-driven and traditional CPS models to identify differences in block types and their frequency. The goal of this analysis is to reveal trends and patterns in block usage that distinguish the design of AI-driven control systems from traditional ones. To answer the research question, we consider all ($8$) systems in our benchmark, with both their AI-driven and traditional models, totaling $16$ models in our study. To ensure that our analysis captures design trends central to system logic and core functionality, we identify, from Simulink Block Libraries~\cite{MathWorks2024}, ``relevant" block types, those directly contributing to control logic, data processing, and decision-making. These relevant blocks are organized by their categories. Table~\ref{tab:matlablibrary} provides an overview of these categories along with examples of representative block types, including Logic Operations (e.g., relational operations) for conditional decision-making as well as Continuous and Discrete elements (e.g., integrator, PID controller) that are essential for managing system dynamics. Blocks outside this set, deemed “irrelevant” to our analysis, are excluded; these include signal checking (e.g., vector checks), visualization (e.g., Scope), static checks, debugging tools, and miscellaneous formatting elements (e.g., Mux, Demux). Note that this filtering process was informed by domain expertise to ensure accurate comparison. The complete list of relevant block types used in the model analysis and examples of irrelevant block types are included in the replication package~\cite{figshare2024} for reference. For each system model, we first identify and isolate the relevant blocks according to our catalog, excluding any blocks deemed irrelevant. Next, we count the presence of each block type within the filtered set. To facilitate this, we utilize MATLAB’s \emph{find\_system}~\cite{MathWorksfindsystem} function, which searches for blocks by type and reports the count for each. We configure the function parameters as follows, with \emph{LookUnderMasks} set to \emph{'all'}, \emph{FollowLinks} set to \emph{'on'}, and \emph{MatchFilter} configured to \emph{Simulink.match.allVariants}. These settings ensure that all blocks are included, even those located within subsystems, variants, and masked blocks. we then calculate the \emph{Difference} for each block type by comparing the occurrence of each block between the AI-driven and traditional model versions. This metric is defined as:
\begin{equation}
    \text{Difference} = AI\_blocks - Traditional\_blocks
\end{equation}

\emph{Difference} indicates the relative occurrence of each block type in AI-driven models compared to traditional models. In this formula, \emph{AI\_blocks} refers to the count of a specific type of atomic block within the AI-driven model, while \emph{Traditional\_blocks} refers to the count of the same block type within the traditional model. We examine the \emph{Difference} values over $8$ categories and we identify those that are more prevalent in either AI-driven or traditional model constructs. For instance, a positive \emph{Difference} value indicates a higher usage of that block type in the AI-driven model, while a negative value indicates a greater usage in the traditional model. 

\begin{figure}[t]
\centerline{\includegraphics[width=0.5\textwidth]{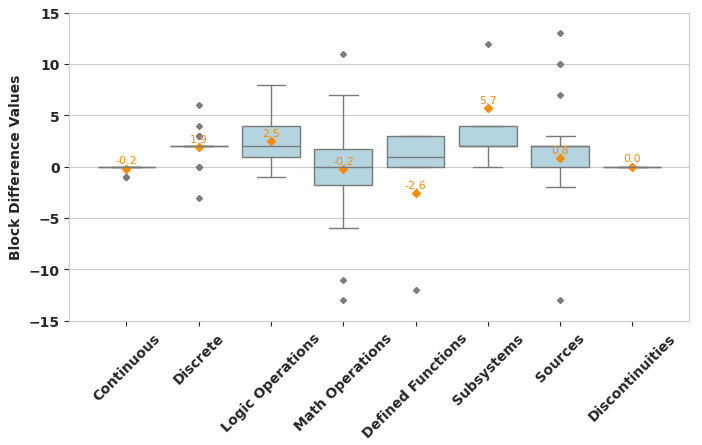}}
\caption{Category-Wise Atomic Block Differences between AI-Driven and Traditional CPS models.}
\label{fig:rq1}
\end{figure}

\textbf{Results.} Figure~\ref{fig:rq1} presents the distribution of \textit{Difference} values between AI-driven and traditional CPS models across the relevant categories. Each box plot highlights the average \textit{Difference} value with a diamond marker. The results reveal significant architectural shifts in block-type usage between the two model constructs. Specifically, AI-driven models show a slightly decreased reliance on \textit{Continuous} blocks (averaging $-0.2$), suggesting a move away from continuous-time control components that have characterized CPS architectures for decades. Traditional control systems rely on continuous-time blocks to enable real-time responses to environmental changes in highly dynamic systems, while maintaining stability. However, continuous-time blocks may not scale as system complexity, size, or capability of the control system increase, often to handle more inputs, outputs, or greater functionality. This difficulty in scaling motivates engineers to shift toward discrete-time control to ensure greater scalability. In contrast, the results show that our AI-driven models make increased use of \textit{Discrete} and \textit{Logic Operations} blocks (averaging $1.9$ and $2.5$ more blocks, respectively) compared to the traditional models, indicating a shift toward discrete-time processing and complex logical decision-making. This trend aligns with AI-driven models’ need to handle asynchronous processes and more adapted decisions. Discrete blocks scale more easily as they are typically implemented through software, modular design, and digital communication networks. However, real-time response in discrete control is limited which may introduce delays compared to continuous-time control. Moreover, the reliance on \textit{Ports \& Subsystems} ($5.7$) category blocks is more pronounced in AI-driven models, likely as a means to compensate for the reduced continuous-time processing. This category allows real-time decision-making that can replace some of the continuous-time functionalities.
The reduced usage of \textit{User-Defined Functions} ($-2.6$) in AI-driven models suggests a preference for prebuilt or embedded functions, contrasting with traditional models' reliance on custom functions to meet specific control needs. \textit{Math Operations} and \textit{Discontinuities}, on the other hand, are used at comparable levels, as both serve foundational roles across the model architectures. These findings suggest that AI-driven CPS models are increasingly structured around modular, discrete, and logic-intensive designs. This modularity facilitates complex control strategies and adaptability, but it also introduces additional layers of dependencies and interactions between components, increasing structural complexity. 

\begin{tcolorbox}[boxsep=0pt,left=3pt,right=3pt,colback=white]
\textbf{RQ1:} The transition from traditional to AI-driven CPS models introduces an evolution in CPS architecture, marked by a reduction in continuous dynamics and an increased reliance on discrete, logic-driven, and modular design. This shift improves adaptability, though with some trade-off of real-time responsiveness, aligning model design with the demands of advanced, AI-integrated environments.
\end{tcolorbox}

\subsection{RQ2}
\begin{figure*}[t]
\centerline{\includegraphics[width=1\textwidth]{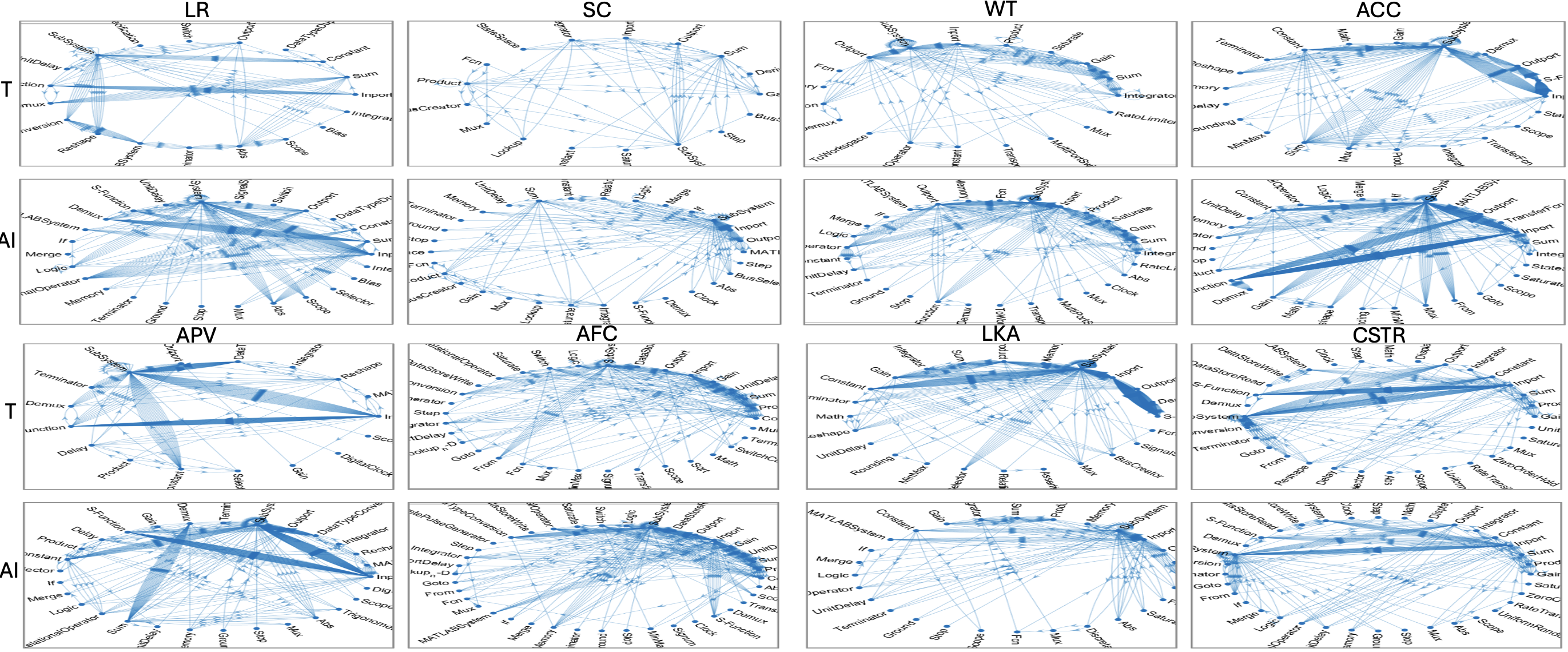}}
\caption{Flow Graph of our Case Study Systems for Traditional (T) and AI-Driven (AI) Models.}
\label{fig:cfgs}
\end{figure*}
To address RQ2, For each model, we generate a customized flow graph as described in Section~\ref{sec:step3}. To ensure fairness, we focus exclusively on relevant connections involving at least one block identified as relevant in RQ1, counting edges where either the source or destination is classified as relevant. We develop a specialized algorithm to generate flow graphs by tracing each model according to our specified criteria, ensuring that the resulting graphs accurately represent the control pathways central to the system's core operation. The algorithm generates a set of $16$ flow graphs, one for each model across $8$ systems described in Table~\ref{tab:cpsmodels}. For each AI-driven and traditional model flow-graph pair, we conduct a comparative analysis to identify connection differences and draw insights into their potential implications. To quantitatively assess the control flow characteristics, we calculate three metrics for each model: block count, connection count, and hierarchical depth (defined in Section~\ref{sec:step3}), for both total and relevant connections. Higher connection counts and hierarchical depth, for example, suggest increased decision branching, which requires more computational time and potentially introduces instability. We examine how AI integration shifts control flow and affects overall complexity in CPS models.

\textbf{Results.} Figure~\ref{fig:cfgs} presents the $16$ flow graphs generated for each model of our case study systems. Table~\ref{tab:structuralresult} provides the quantitative analysis results for these models. Each row in the table corresponds to one system from our benchmark set, while the columns detail the flow dynamics metrics for each system and model, including \emph{Total BC} (i.e., the number of blocks in the entire model), \emph{Relevant BC} (i.e., the number of relevant blocks), \emph{Total CC} (i.e., the number of edges or connections between all blocks), \emph{relevant CC} (the number of connections between relevant blocks and other blocks), and \emph{HD} (i.e., the longest path from the root node to a leaf node within the model graph). The flow graphs illustrate greater inter-connections in AI-driven models compared to the traditional models across most of our systems ($6$ out of $8$). Table~\ref{tab:structuralresult} also reveals higher block and connection counts and deeper hierarchical structures in AI-driven models for the same cases illustrated in the figure. Specifically, traditional models have a lower number of atomic blocks compared to AI models in $6$ out of $8$ systems (e.g., LR, SC, WT, APV, ACC, and AFC), with an average block count of $288.75$ and $349.13$, and average relevant block count of  $116.5$ and $141.5$, for Traditional and AI-driven models, respectively. These AI-driven systems demonstrate higher inter-connectivity between nodes than their traditional counterparts, as shown in both the system flow graphs and the connection counts presented in the table (see highlighted values in Table~\ref{tab:structuralresult}). For Traditional and AI-driven models, respectively, the average total \emph{CC} are $306$ and $350.13$, while the relevant \emph{CC} are $274.25$ and $312$. This suggests that AI-driven models incorporate more complex decision-making processes and more data flows across blocks. Much of the added connectivity are a result of additional feedback loops or internal states unique to AI models, which can introduce more dynamic responses but also higher potential for complexity and a higher likelihood of errors-prone behaviors. The hierarchical depth results, show longer and more elaborate paths in AI-driven models, which exhibit greater depth compared to traditional models in $6$ out of $8$ systems and $21\%$ average increase. A deeper hierarchy reflects additional layers, likely required to accommodate the increased complexity of adaptive control in AI-driven systems. Note that, in order to ensure clarity in our analysis, we exclude the inport and outport blocks along with their direct connections, due to their disproportionately large number compared to other block types, which could obscure insights into the remaining system components and inter-connections. We analyzed separately their counts and connections across models. The results show an average increase of $25.5$ inports and $9.37$ outports in AI-driven models compared to traditional models, with total connection counts combining inports and outports of $1,445$ for AI-driven models and $1,160$ for traditional models, contributing to increased system complexity. Overall, our analysis reveals that AI-driven models of CPS feature more complex decision branching and inter-connectivity compared to traditional models, indicating greater adaptability to changing environment. However, more complex pathways could affect real-time performance, as more complex paths require more computational resources, which may slow down response times. 

\begin{table}[t]
\centering
\caption{Flow Dynamics Metrics: Total and Relevant Block Count (BC), Total and Relevant Connection Count (CC), and Hierarchical Depth (HD) for Traditional (T) and AI-Driven (AI) CPS Models}
\label{tab:structuralresult}
\centering
\begin{tabular}{|l|l|c|c|c|c|c|}
\hline
\textbf{System} & \textbf{Model} & \textbf{Total} & \textbf{Relevant} & \textbf{Total} & \textbf{Relevant} & \textbf{HD} \\
 &  & \textbf{BC} & \textbf{BC} & \textbf{CC} & \textbf{CC} & \\
\hline
ACC & T   & \cellcolor{yellow}390  & \cellcolor{yellow}142 & \cellcolor{yellow}365 & \cellcolor{yellow}334   & 7 \\
    & AI  & \cellcolor{yellow}591  & \cellcolor{yellow}211 & \cellcolor{yellow}525 & \cellcolor{yellow}475  & 9 \\
\hline
AFC & T   & \cellcolor{yellow}302  & \cellcolor{yellow}154 & \cellcolor{yellow}337 & \cellcolor{yellow}295   & 7 \\
    & AI  & \cellcolor{yellow}426  & \cellcolor{yellow}191 & \cellcolor{yellow}460 & \cellcolor{yellow}407  & 8 \\
\hline
LKA & T   & 604 & 219 & 608 & 546  & 9 \\
    & AI  & 210 & 86 & 208 & 189  & 7 \\
\hline
LR  & T   & \cellcolor{yellow}168  & \cellcolor{yellow}76 & \cellcolor{yellow}198 & \cellcolor{yellow}170   & 7 \\
    & AI  & \cellcolor{yellow}252  & \cellcolor{yellow}107 & \cellcolor{yellow}289   & \cellcolor{yellow}243 & 7 \\
\hline
SC  & T   & \cellcolor{yellow}61   & \cellcolor{yellow}35 & \cellcolor{yellow}82 & \cellcolor{yellow}73  & 5 \\
    & AI  & \cellcolor{yellow}215  & \cellcolor{yellow}89 & \cellcolor{yellow}210 & \cellcolor{yellow}184 & 8 \\
\hline
WT  & T   & \cellcolor{yellow}175  & \cellcolor{yellow}88 & \cellcolor{yellow}212 & \cellcolor{yellow}203 & 6 \\
    & AI  & \cellcolor{yellow}350  & \cellcolor{yellow}159 & \cellcolor{yellow}367 & \cellcolor{yellow}343 & 8 \\
\hline
APV & T   & \cellcolor{yellow}260  & \cellcolor{yellow}89 & \cellcolor{yellow}282 & \cellcolor{yellow}247 & 5 \\
    & AI  & \cellcolor{yellow}458  & \cellcolor{yellow}173 & \cellcolor{yellow}468 & \cellcolor{yellow}411 & 8 \\
\hline
CSTR & T  & 350 & 129 & 364 & 326  & 5 \\
     & AI & 291 & 116  & 274 & 244   & 7 \\
\hline
\hline
\hline
\textbf{Average} & T & \textbf{288.75} & \textbf{116.5} & \textbf{306} & \textbf{274.25} & \textbf{6.38} \\
      & AI & \textbf{349.13} & \textbf{141.5} & \textbf{350.13} & \textbf{312} & \textbf{7.75} \\
\hline\hline
\hline
\textbf{\% Diff} & - & \textbf{+29.2} & \textbf{+25.7} & \textbf{+21.1} & \textbf{+20.5} & \textbf{+21.0} \\
\hline
\end{tabular}
\end{table}

\begin{tcolorbox}[boxsep=0pt,left=3pt,right=3pt,colback=white]
\textbf{RQ2:} AI-driven models exhibit more complex dynamic flows compared to traditional models, with \textbf{25.7\%} average increase in the relevant blocks and \textbf{20.5\%} average increase in connectivity, to support adaptability to varying conditions, as AI-driven models integrate more feedback loops and decision points. However, this may impact stability and real-time performance, as deeper structures require more computational resources and may slow response times.
\end{tcolorbox}

\subsection{RQ3}
\begin{table*}
\centering
\caption{Requirements of AFC, WT and SC systems formulated in natural language and STL.}
\label{tab:requirements}
\begin{tabular}{p{0.05\textwidth} | p{0.05\textwidth} | p{0.3\textwidth} | p{0.5\textwidth}}
\hline
\textbf{System} & \textbf{ReqID} & \textbf{Description} & \textbf{STL Formula} \\
\hline
AFC & AFC27 & If there’s a ``rise" or ``fall" between $11$ and $50\sec$, then $\mu$ must stay below $\beta$ within $5\sec$.& $G_{[11, 50]} ((\mathit{rise} \lor \mathit{fall}) \to (G_{[1,5]} |\mu| < \beta)), \text{ where } \mathit{rise} = (\theta < 8.8) \land (F_{[0, 0.05]} (\theta > 40.0)), \mathit{fall} = (\theta > 40.0) \land (F_{[0, 0.05]} (\theta < 8.8))$, $\beta = 0.008$, \\
 & AFC29 & From $11$ to $50$ $\sec$, $\mu$ should stay below \( \gamma \). & $G_{[11, 50]} |\mu| < \gamma$, where $\gamma = 0.008$ \\
 & AFC33 & From $11$ to $50\sec$, $\mu$ should stay below \( \gamma \). & $G_{[11, 50]} |\mu| < \gamma$, where $\gamma = 0.007$ \\
\hline
WT  & WT1 & From $30$ to $630\sec$, \( \theta \) must remain below 14.2. & $G_{[30, 630]} \theta \leq 14.2$ \\
 & WT2 & From $30$ to $630\sec$, the torque must be within $21,000Nm$ and $47,500Nm$. & $G_{[30, 630]} 21000 \leq M_{g, d} \leq 47500$ \\
 & WT3 & From $30$ to $630\sec$, \( \Omega \) must remain below $14.3$. & $G_{[30, 630]} \Omega \leq 14.3$ \\
 & WT4 & The absolute difference between \( \theta \) and \( \theta_d \) should not exceed $1.6$ for more than $5sec$. & $G_{[30, 630]} F_{[0,5]} |\theta - \theta_d| \leq 1.6$ \\
\hline
SC  & SC & The pressure should stay within $87$ to $87.5Pa$. & $G_{[30,35]} (87\leq \mathit{pressure} \land \mathit{pressure}\leq 87.5)$ \\
\hline
\end{tabular}
\end{table*}

To address \textit{RQ3}, we assess how AI integration affects the effectiveness of existing CPS verification processes, specifically using \emph{S-TaLiRo} to detect faults in CPS models that lead to requirement violations. \emph{S-TaLiRo} is a MATLAB-based falsification tool widely applied in verifying continuous and hybrid dynamic systems using linear-time temporal logic. It performs automated testing by generating test cases through stochastic optimization techniques, aiming to find input signals that steer system behaviors to violating specified temporal logic requirements. In our experiments, we configure \emph{S-TaLiRo} to falsify both AI-driven and traditional CPS models of three systems in our Benchmark. Each system comes with a set of functional requirements, specified by the ARCH competition~\cite{ernst2022arch}. Requirements for each system are formulated in Signal Temporal Logic (STL)~\cite{annpureddy2011staliro}, a formalism that precisely defines temporal and logical constraints over system signals. Table~\ref{tab:requirements} shows the STL formula associated to each system requirement. For each model, we create configuration files specifying the requirements, input ranges, simulation time, and the number of control points for the generated signals. Using these inputs, \emph{S-TaLiRo} executes three primary steps: (1) generating an input signal within the defined parameters, (2) simulating the model to produce an output trace based on the input, and (3) checking the output trace against STL requirements to detect any violations. The tool’s output reports whether a fault-finding (violating) trace was found. In a first experiment (\textit{EXP-I}), we select three representative models from our benchmark, i.e., AFC, WT, and SC, to evaluate the effectiveness of \emph{S-TaLiRo} in detecting requirement violations across both AI-driven and traditional CPS models, evaluating a total of $8$ requirements. We delegated these representative models due to the computational expense of running all models and their requirements through $30$ executions with a maximum of $300$ iterations per execution. We set the optimization algorithm as \emph{Simulated Annealing} (SA) based on its extensive usage in prior \emph{S-TaLiRo} studies~\cite{gaaloul2021combining}. The AI-driven models are configured with the \emph{Deep Deterministic Policy Gradient} (DDPG) policy. This experiment serves to assess the fault-detection capabilities of \emph{S-TaLiRo} when applied to AI-driven models compared to traditional models. In a second experiment \textit{EXP-II}, we run \emph{S-TaLiRo} on the AI-driven SC model where we analyze its performance under four distinct policies: \emph{DDPG}, \emph{Twin-Delayed Deep Deterministic Policy Gradient} (TD3), \emph{Actor-Critic} (A2C), and \emph{Proximal Policy Optimization} (PPO). We evaluate the effectiveness of \emph{S-TaLiRo} in detecting faults across various AI policies, comparing its performance to the traditional model of SC.
\begin{table}[t]
\centering
\caption{(EXP-II) Fault Detection Results of \emph{S-TaLiRo} in SC Model: Model version and AI Policy; Number of Executions w/ Violations; Number of Falsified Requirements, Average execution time in seconds.}
\label{tab:exp2_results}
\begin{tabular}{|l|c|c|||c|}
\hline
\textbf{Model/Policy} & \textbf{\#Violated Exec.} & \textbf{Avg. time} & \textbf{\# Fals. Requirements} \\ 
 & \textbf{(SC)} & \textbf{(SC)} & \textbf{(All models)} \\
\hline
Traditional & 30 & 0.2 & 8/8\\ 
\hline
A2C & 29 & 70.2 & 6/8 \\ 
DDPG & 26 & 59.4 & 6/8 \\ 
TD3 & 25 & 80.4 & 6/8 \\ 
PPO & 24 & 85.6 & 6/8 \\ 
\hline
\textbf{AI Avg} & \textbf{26} & \textbf{73.9} & \textbf{6/8}\\
\hline
\end{tabular}
\end{table}

\textbf{Results.} The results of \textit{EXP-I} indicate that \emph{S-TaLiRo} detected requirement violations in traditional CPS models with high consistency, identifying faults in an average of $26.25$ executions out of $30$ and successfully falsifying $7$ out of $8$ requirements. However, for AI-driven CPS models, \emph{S-TaLiRo} detected violations in an average of only $18.25$ executions, covering just $5$ requirements.  presents the fault-detection results of \emph{S-TaLiRo} for the SC system under both traditional model and AI-driven control policies. The results are summarized in terms of the number of violated executions out of 30, the number of falsified requirements out of 8, and the average time required to falsify the system requirements. The last row of the table provides the average values across all four AI policies. The results of \textit{EXP-II} show that \emph{S-TaLiRo} detected violations across all $30$ executions for the traditional SC model. However, for the AI-driven SC model, the tool’s fault-detection success varied across policies: $29$ executions for A2C, $26$ for DDPG, $25$ for TD3, and $24$ for PPO. The average time to identify a violation differed significantly, with traditional SC requiring just $0.2$ seconds, while the AI policies took significantly longer: $70.2s$ for A2C, $59.4s$ for DDPG, $80.4s$ for TD3, and $85.6s$ for PPO. These results highlight that while \emph{S-TaLiRo} effectively detects violations in the traditional SC model with minimal computational time, its performance in AI-driven models is lower, with longer detection times and varying success rates across different policies. This suggests that the complexity introduced by AI impacts \emph{S-TaLiRo}’s fault-detection effectiveness, which varies across AI policies, indicating the need for adapted verification strategies to handle different AI configurations.
Overall, traditional models, with simpler and more deterministic paths, offer enhanced stability and predictability, which simplifies verification and validation processes. While AI-driven models may struggle with real-time responsiveness due to the complexity of their decision-making processes, traditional controllers are generally better suited to meet real-time constraints. This trade-off between adaptability in AI-driven models and stability in traditional counterparts highlights the need for additional verification measures or adaptations in AI-driven systems to ensure reliable performance across all expected conditions.

\begin{tcolorbox}[boxsep=0pt,left=3pt,right=3pt,colback=white]
\textbf{RQ3:} Out of the 8 requirements, \emph{S-TaLiRo} successfully falsified a greater number of requirements in traditional models (\textbf{7}) compared to AI-driven models (\textbf{5}). Moreover, \emph{S-TaLiRo} required significantly more computational time for AI-driven models, with an average execution time of \textbf{73.9} seconds, compared to just \textbf{0.2} seconds for the traditional model. This highlights that adaptability of AI-driven model entails trade-offs with reliability. Therefore, there is need for more adapted verification approaches to cope with the complexity and unpredictability introduced by AI in CPS.
\end{tcolorbox}

\section{Threats to Validity}
\label{sec:treats}
In this section, we outline the potential threats to the validity of our study and the steps taken to mitigate them.

\textit{Internal Threats:} In our analysis, we filtered out blocks deemed irrelevant to control logic, focusing on relevant components only. Although this approach minimizes noise, it may slightly risk omitting elements specifying system dynamics. To mitigate this, we conducted a sanity check to ensure that irrelevant blocks do not impact centralized control block outcomes. The results across all RQs are aligned which consolidates the validity of our comparative analysis.

Due to computational constraints, our evaluation of \emph{S-TaLiRo} for assessing the verification impact of CPS transformations was conducted on a representative subset of models in our benchmark. While this approach may not fully generalize across all case study systems due to the variability in \emph{S-TaLiRo}'s effectiveness across different CPS landscapes, our comparison in this study remains model-specific rather than tool-specific. We evaluated \emph{S-TaLiRo}'s effectiveness on both AI-driven and traditional models of the same system, with results demonstrating its superior performance and higher efficiency when applied to the traditional model. Further, \emph{S-TaLiRo} has consistently demonstrated success in prior research, including the ARCH competition, which proves its suitability for complex CPS models.

\textit{External Threats:} The genralizability of our study subjects may be impacted for not capturing the full diversity of CPS architectures. To mitigate this, we considered systems from various domains and of varying sizes to ensure representativeness across different applications. Furthermore, to explore the AI-driven behavior space, we involved an experiment where we evaluate multiple types of reinforcement learning controllers using different agent configurations.

\section{Related Works}
\label{sec:related}
\textit{AI Integration in CPS.}
Recent studies~\cite{bucsoniu2018reinforcement, song2022cyber, mauludin2024simulation} investigate AI integration in CPS, particularly using reinforcement learning and neural networks, and discuss challenges and benefits in applications like autonomous vehicles and industrial automation. Schoning et al.~\cite{schöning2023aiintheloopimpacthmi, schoning2022ai, electronics12163489} explore enhancing control design using lightweight ANN architectures in closed-loop control systems (CLCS). They highlight ANN-based controllers that replace traditional control systems improve adaptability in complex environments, but introduce increased complexity, recommending fewer trainable parameters to mitigate computational demands. Busoniu et al.~\cite{bucsoniu2018reinforcement} address the complexities that DRL and AI-driven approaches bring compared to traditional control methods. They note that DNNs excel in modeling complex, nonlinear systems but at the cost of higher computational demands, potential instability, and overfitting risks, especially in dynamic systems, that is guaranteed by traditional methods.

\textit{Verification Practices of AI-enabled CPS.} Studies on AI-enabled CPS design and verification~\cite{radanliev2021artificial} highlight performance improvement but face significant challenges with verifying data-driven neural networks. Limited optimization algorithms in current verification tools impact broader adoption~\cite{zhang2022falsifai, 10144351}. Xuan et al.\cite{xie2023mosaic} propose an abstract model-guided falsification approach using combined local and global search for improved exploration-exploitation balance; however, they lack comparisons with leading tools like S-TaLiRo. Schoning et al.\cite{schoning2022ai, electronics12163489} emphasize that, due to reliability limitations, fully AI-based controllers are not yet feasible for safety-critical applications, suggesting a hybrid approach to enable controlled assessments of AI’s risks and benefits. Other studies~\cite{du2020marble,wang2018automatically} apply abstraction and robustness-guided falsification to RNNs, but struggle with high dimensionality. Reachability analysis techniques~\cite{huang2019reachnn,dreossi2016parallelotope,kong2015dreach,xiang2018reachability,dutta2019reachability,bogomolov2019juliareach} address high-dimensional reachability challenges in AINNCS by implementing dimension reduction to manage the "wrapping effect."

\section{Conclusion}
\label{sec:conclusion}
In this paper, we presented a multi-method approach that combines architectural analysis with a systematic evaluation of standard verification practices to investigate how AI integration in CPS reshapes system architectures to support adaptability in increasingly complex and dynamic environments. Our study provided insights into the architectural shifts required to accommodate AI, addressing emerging verification challenges and implications for safe and reliable system operation. We identified fundamental structural differences between AI-driven and traditional CPS models, including increased size, inter-connectivity, and dynamic responsiveness in AI-driven models. Our results show that AI-driven models exhibit greater size, inter-connectivity, dependencies, and dynamic responsiveness. While these features enhance flexibility, they also introduce verification challenges in fault detection, computational efficiency, and robustness, highlighting the need for adaptive frameworks and tools tailored to AI-driven architectures. Future work will extend this approach to a broader set of case studies and analysis, aiming for a large-scale empirical evaluation comparing advanced verification tools across diverse AI-driven and traditional CPS models.


\section{Data Availability}
Our replication package, available at \cite{figshare2024}, includes the implementation of our study, evaluation data and scripts for generating the presented graphs and results. The package ensures full reproducibility of our findings. 


\bibliographystyle{IEEEtran}
\bibliography{reference}

\vspace{12pt}

\end{document}